\documentclass[amsmath,amssymb,twocolumn]{revtex4}
\usepackage{graphicx}% Include figure files
\usepackage{dcolumn}% Align table columns on decimal point
\usepackage{bm}% bold math
\begin{document}

\title{Traits and Characteristics of Interacting Dirac fermions \break
in Monolayer and
Bilayer Graphene\footnote{\bf Invited Review: Solid State Communications, \break
Special Issue on Graphene}}
\author{Tapash Chakraborty$^\ddag$}
\affiliation{Department of Physics and Astronomy,
University of Manitoba, Winnipeg, Canada R3T 2N2}
\author{Vadim M. Apalkov}
\affiliation{Department of Physics and Astronomy, Georgia State University,
Atlanta, Georgia 30303, USA}

\date{\today}
\begin{abstract}
The relativistic-like behavior of electrons in graphene significantly influences 
the interaction properties of these electrons in a quantizing magnetic field, 
resulting in more stable fractional quantum Hall effect states as compared to those
in conventional (non-relativistic) semiconductor systems. In bilayer graphene 
the interaction strength can be controlled by a bias voltage and by the orientation
of the magnetic field. The finite bias voltage between the graphene monolayers 
can in fact, enhance the interaction strength in a given Landau level. As a function 
of the bias voltage, a graphene bilayer system shows transitions from a state 
with weak electron-electron interactions to a state with strong interactions. 
Interestingly, the in-plane component of a tilted magnetic field can also alter 
the interaction strength in bilayer graphene. We also discuss the nature of the
Pfaffian state in bilayer graphene and demonstrate that the stability of this state
can be greatly enhanced by applying an in-plane magnetic field.
\end{abstract}
%\pacs{}
\maketitle

\section{Introduction} 

Graphene is a monolayer of carbon atoms, which has a two-dimensional (2D) honeycomb 
crystal structure. The unique feature of graphene is that the single-electron 
low-energy dispersion has the relativistic massless form commonly attributed to the 
Dirac fermions, and the corresponding electron wave functions have a chiral nature 
\cite{abergeletal,ando_07,book_kats,book_raza}. The electronic band 
structure, first derived by Wallace in 1947 \cite{wallace} has two valleys at two 
inequivalent corners, $K=(2\pi/a)(\frac13,\frac1{\sqrt3})$ and $K^{\prime}=(2\pi/a)(\frac23,0)$, 
of the Brillouin zone, where $a=0.246$ nm is the lattice constant. The low-energy dispersion 
at the valleys $\xi = 1 $ ($K$-valley) and $\xi = -1$ ($K^{\prime}$-valley) is determined 
by the following relativistic massless Hamiltonian 
\cite{abergeletal,ando_07,book_kats,book_raza,geim_rise,netoetal}
\begin{equation}
{\cal H}^{}_{\xi} = \xi v^{}_{\rm F}\left( 
\begin{array}{cc}
    0 & p^{}_{-} \\
    p^{}_{+} & 0  
 \end{array} 
\right),
\label{H}
\end{equation}
where $p^{}_- = p^{}_x- {\rm i} p^{}_y$, $p^{}_+ = p^{}_x + {\rm i} p^{}_y$,  
and $\vec{p}$ is the two-dimensional momentum of an electron. Here $v^{}_{\rm F} 
\approx 10^6$ m/s is the Fermi velocity, which is related to the hopping 
integral between the nearest neighbor sites.  The honeycomb lattice of graphene
consists of two sublattices A and B and the two component wave functions corresponding to 
the Hamiltonian (\ref{H})  can be expressed as $(\psi^{}_{A}, \psi^{}_B)^{T} $ for valley $K$ 
and $(\psi^{}_B, \psi^{}_A)^{T}$ for valley $K^{\prime}$, where $\psi^{}_A$ 
and $\psi^{}_B$ are wave functions of sublattices $A$ and $B$, respectively. 

The two components of the wave function correspond to the quantum mechanical amplitudes 
of finding the `Dirac fermion' on one of the two sublattices. This sublattice degree 
of freedom is often referred to as {\it pseudospin} of Dirac fermions in graphene. It is 
directed along the direction of motion of the Dirac fermion in the conduction band, and 
opposite to the motion in the valence band. In other words, particles in graphene have opposite
{\it chirality} in the $K$ and $K^{\prime}$ valleys. The electron and hole wave functions
are eigenfunctions of the helicity (chirality) operator \cite{book_kats}. Physically,
a certain direction of the pseudospin in the graphene plane corresponds to a rotation
of the relative phases of the two components of the spinor eigenstates along that
direction of motion of the Dirac fermion. The sublattice pseudospin chirality of
Dirac fermions does not allow perfect backscattering (between states of opposite
momentum and opposite pseudospin) that has important consequences on the physical
characteristics of graphene \cite{book_kats,regan}. 

The single-electron states obtained from the Hamiltonian (\ref{H}) has 
a {\it linear} relativistic dispersion relation of the form 
\begin{equation}
\varepsilon(p)=\pm v^{}_{\rm F} p,
\label{energy}
\end{equation}
where the signs `+' and `-' correspond to the conduction and valence bands, respectively.
Each energy level (\ref{energy}) is four-fold degenerate due to two-fold spin and two-fold 
valley degeneracies. 

\section{Dirac Fermions in Magnetic Fields} 

In a magnetic field applied perpendicular to the graphene plane, the relativistic energy 
dispersion relation (\ref{energy}) brings in very specific form of Landau levels 
of electrons in graphene \cite{mcclure,haering}, and as a consequence the 2D system
displays unconventional quantum Hall effects. The Landau levels of electrons in graphene 
can be found from the Hamiltonian (\ref{H}) by replacing the electron momentum $\vec{p}$ 
with the generalized momentum $\vec{\pi} = \vec{p} + e\vec{A}/c$. Here $\vec{A}$ is the 
vector potential. Then the Hamiltonian of an electron in a magnetic field perpendicular 
to the graphene monolayer in valley $\xi$ takes the form 
\begin{equation}
{\cal H}^{}_{\xi } = \xi v^{}_{\rm F} \left( 
\begin{array}{cc}
    0 & \pi^{}_{-}  \\
    \pi^{}_{+} & 0  
 \end{array} 
\right).
\label{Hm}
\end{equation}
The eigenfunctions of the Hamiltonian (\ref{Hm}) can be expressed in terms 
of the conventional Landau level wave functions, $\phi^{}_{n,m}$, for a particle 
obeying the parabolic dispersion relation with the Landau index $n$ and intra-Landau 
index $m$, which depends on the choice of the gauge. For example, in the Landau 
gauge ($A^{}_x =0$ and $A^{}_y= Bx$) the index $m$ is the $y$-component of the 
momentum, while in the symmetric gauge ($\vec{A} = \frac12\vec{B}\times \vec{r}$) 
the index $m$ is the $z$-component of electron angular momentum. For these 
wave functions, $\phi^{}_{n,m}$, the operators $\pi^{}_{+}$ and $\pi^{}_{-}$ are 
the raising and lowering operators, respectively. 

The Landau eigenfunctions of the Hamiltonian (\ref{Hm}) are then written in the form
\begin{equation}
\Psi^{K}_{n,m} = \left(\begin{array}{c}
 \psi^{}_A \\
   \psi^{}_B
\end{array}  
 \right) = C^{}_n
\left( \begin{array}{c}
 {\rm sgn}(n) {\rm i}^{|n|-1}\phi^{}_{|n|-1,m} \\
    {\rm i}^{|n|} \phi^{}_{|n|,m}   
\end{array}  
 \right),
\label{f1}
\end{equation}
for valley $K$ ($\xi = 1$) and 
\begin{equation}
\Psi^{K^{\prime}}_{n,m} =\left(\begin{array}{c}
 \psi^{}_B \\
   \psi^{}_A
\end{array}  
 \right)=C^{}_n
\left( \begin{array}{c}
 {\rm sgn}(n) {\rm i}^{|n|-1} \phi^{}_{|n|-1,m} \\
    {\rm i}^{|n|} \phi^{}_{|n|,m}   
\end{array}  
 \right),
\label{f2}
\end{equation}
for valley $K^{\prime}$ ($\xi=-1$). Here $C^{}_n=1 $ for $n=0$ and 
$C^{}_n=1/\sqrt2$ for $n\neq 0$ and 
\begin{equation}
\mbox{sgn}(n) = \left\{ \begin{array}{cc}
0 &  (n=0)  \\
1 &  (n>0)  \\
-1 &  (n<0).
\end{array} \right.,
\end{equation} 
where positive and negative values of $n$ correspond to the conduction and 
valence bands, respectively. The corresponding Landau energy spectrum takes 
the form 
\begin{equation}
\varepsilon^{}_n=\hbar\omega^{}_B\mbox{sgn}(n)\sqrt{|n|},
\label{landau}
\end{equation}
where $\omega^{}_B=\sqrt2 v^{}_{\rm F}/\ell^{}_0$ and $\ell^{}_0=\sqrt{\hbar/e B}$ 
is the magnetic length. 

\begin{figure}
\begin{center}\includegraphics[width=8cm]{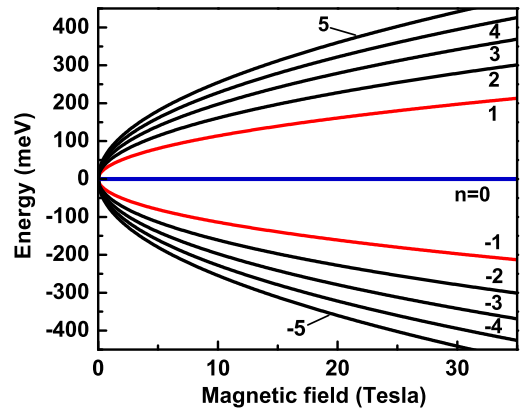}
\end{center}
\caption{\label{Fig_LL}
The Landau levels as a function of the perpendicular magnetic field for a graphene 
monolayer. Numbers next to the lines are the Landau level indices. The Landau levels 
with positive energies (positive index $n$) and negative energies (negative index 
$n$) correspond to the conduction and valence bands, respectively. The fractional
quantum Hall effect (FQHE) \cite{FQHE_book} that is discussed below, can be observed 
only in the Landau levels shown by the red and blue lines ($n=\pm 1$ and 0). The 
strongest electron-electron interactions and correspondingly the more stable FQHE states 
are realized in the Landau levels shown by red lines.  
}
\end{figure}

The specific feature of the Landau levels in graphene is their square-root 
dependence on both the magnetic field $B$ and the Landau level index $n$. This 
behavior is different from that in conventional ({\it non-relativistic}) 
semiconductor 2D system with the parabolic dispersion relation, for which the 
energy spectrum has a linear dependence on both the magnetic field and the 
Landau level index, $\varepsilon^{}_n=\hbar\omega^{}_B (n+1/2)$. In Fig.\ \ref{Fig_LL} 
the Landau levels in graphene are shown as a function of the perpendicular magnetic 
field, where the positive and negative Landau level indices ($n$) correspond to the 
conduction and valence bands, respectively. The Dirac nature of the electron dynamics 
in graphene and the unique behavior of Landau levels in a graphene monolayer,
were experimentally confirmed by observation of quantum Hall plateaus at filling 
factors $\nu=4\left(n+\frac12\right)$ \cite{Novoselov_2005,Zhang_2005}. 

%-------------------------------------------------------------------- 
%-----------------------------------------------------------

The interaction properties of electrons within a single Landau level, 
i.e., disregarding the mixture of Landau levels due to electron-electron interactions, 
are entirely determined by the pseudopotentials $V^{(n)}_m$ proposed by Haldane 
\cite{haldane1,haldane2} which are defined as the energy of two electrons with 
relative angular momentum $m$. They are determined by the structure of the wave 
functions of the corresponding Landau level and for the $n$-th Landau level can be 
evaluated from the following expression \cite{haldane_87}
\begin{equation}
V_m^{(n)} = \int_0^{\infty } \frac{dq}{2\pi} q V(q)
\left[F^{}_n(q) \right]^2 L^{}_m (q^2)
 {\rm e}^{-q^2},
\label{Vm}
\end{equation}
where $L^{}_m(x)$ are the Laguerre polynomials, $V(q) = 2\pi {\rm e}^2/(\kappa 
\ell^{}_0 q)$ is the Coulomb interaction in the momentum space, $\kappa$ is the 
dielectric constant, and $F^{}_n(q)$ is the form factor of the $n$-th Landau level. 
In what follows, all pseudopotentials are given in units of the Coulomb energy, 
$\varepsilon^{}_C = e^2 /\kappa\ell^{}_0$. Equation\ (\ref{Vm}) is valid for 
all types of electron systems (non-relativistic, monolayer and bilayer graphene, etc.) 
with well defined 2D Landau levels. The difference between these systems is 
in the expression of the form factors, $ F^{}_n(q)$. In a non-relativistic 
system, for which the Landau level wave functions are $\phi^{}_{n,m}$, the form 
factors are obtained from $F^{}_n(q) =  L^{}_n\left( q^2/2\right)$. 
In the case of  graphene, the $n$-th Landau level wave functions are given by 
Eqs.\ (\ref{f1})-(\ref{f2}), which results in the following expressions 
\cite{Apalkov_2006,Goerbig_06} for the corresponding form factors 
\begin{eqnarray}
& & F^{}_0(q)=L^{}_0\left(\frac{q^2}2\right) \label{f0}  \\
& & F^{}_n(q)=\frac12\left[L^{}_n \left(\frac{q^2}2\right) 
  +  L^{}_{n-1}\left(\frac{q^2}2\right)\right].
\label{fn}
\end{eqnarray}
With these form factors the pseudopotentials for graphene are then evaluated 
from Eq.~(\ref{Vm}).  

One of the unique manifestations of electron-electron interactions within a single Landau 
level is the formation of incompressible fractional quantum Hall effect (FQHE) 
states \cite{FQHE_book}, which are characterized by a finite excitation gap, determined 
by the electron-electron interactions. These states are realized at the fractional filling 
of a given Landau level, e.g., at filling factors $\nu = 1/m$, where $m$ is an odd 
integer. The properties of the FQHE states are completely determined by the corresponding 
pseudopotentials $V_m^{(n)}$. The stability of the incompressible FQHE state, i.e., 
the magnitude of the FQHE gap depends on how fast the pseudopotentials decay with 
increasing relative angular momentum. For spin and valley polarized electron systems this 
decay is determined by the ratios $V_1^{(n)}/V_3^{(n)}$ and $V_3^{(n)}/V_5^{(n)}$. 
A more stable FQHE is expected for Landau levels when the ratio of the pseudopotentials
is large. 

In Table \ref{tableVm} the values of the ratios are shown for graphene and for 
non-relativistic systems for two lowest Landau levels with $n=0$ and 1 (only in 
these Landau levels the FQHE can be observed). For the non-relativistic system the 
most stable FQHE is observed for the $n=0$ Landau level, which is supported by the 
data in Table \ref{tableVm}, where $V_1^{(n)}/V_3^{(n)}$ for the non-relativistic 
system is the largest in the $n=0$ Landau level. A different situation occurs for 
the graphene system. Here for the $n=0$ Landau level, the wave functions are 
identical to the non-relativistic $n=0$ Landau level wave functions 
[Eqs.~(\ref{f1})-(\ref{f2})]. Therefore, the properties of the FQHE for the $n=0$ Landau 
levels of a non-relativistic system and graphene are the same if expressed in units 
of $\varepsilon^{}_C $. The wave functions of the $n=1$ Landau level of graphene is 
the mixture of the $n=0$ and $n=1$ non-relativistic wave functions, which results in 
an {\it enhancement} of the electron-electron interaction strength for the  $n=1$ graphene 
Landau level \cite{Apalkov_2006,Goerbig_06}. In this Landau level, the ratio 
$V_1^{(n)}/V_3^{(n)}$ has the largest 
value (Table \ref{tableVm}), which suggests that the gaps of the FQHE states 
should have the largest value in graphene for the $n=1$ Landau level.  

\begin{table}
\begin{center}
  \caption{\label{tableVm} 
  Characteristics of the pseudopotentials for 
  graphene and for conventional electron systems}
  \begin{tabular}{l|c|c} 
      & $V_1^{(n)}/V_3^{(n)}$ & $V_1^{(n)}/V_3^{(n)}$ \\ \hline
   $n=0$ (graphene)           & 1.60 & 1.26 \\ 
   $n=0$ (non-relativistic)   &  &   \\ \hline   
   $n=1$ (graphene)           & 1.68 & 1.33 \\ \hline 
   $n=1$ (non-relativistic)   & 1.32 & 1.36 \\ \hline
  \end{tabular} 
\end{center}  
\end{table}

In theoretical studies the FQHE is often analyzed by numerical diagonalization of 
the Hamiltonian matrix for finite-size electron systems in either the planar (torus) 
or the spherical geometry \cite{FQHE_book}. In the case of the spherical geometry 
\cite{haldane1,haldane2,greiter} the magnetic field is introduced in terms of the 
integer number $2S$ of magnetic fluxes through the sphere in units of the flux 
quantum, where the radius of the sphere $R$ is defined as $R=\sqrt{S}\ell^{}_0$. 
The number of available states in a sphere, which corresponds to the states of a 
single Landau level in planar geometry, is $(2S+1)$. For a given number of electrons 
$N^{}_e$ the parameter $S$ determines the filling factor of the Landau level. For example, 
for the filling factor $\nu = 1/m$ it is $S = (m/2)(N^{}_e-1)$. In the case
of the many-electron system the lowest eigenvalues of the interaction Hamiltonian 
matrix determine the nature of the FQHE state and the corresponding neutral excitation 
gap \cite{fano}. The numerical results obtained for a finite size system show that 
the FQHE excitation gap in graphene is the largest for the $n=1$ Landau level 
\cite{Apalkov_2006,Apalkov_2007,Toke_2006,Shibata_2009}. As an example, for $N^{}_e=8$ 
electrons the excitation gap is  $0.083 \varepsilon^{}_C$ for the $n=0$ Landau level 
and $0.094 \varepsilon^{}_C$ for the $n=1$ Landau level. This behavior is consistent 
with the properties of the pseudopotentials shown in  Table \ref{tableVm}.

In Fig.\ \ref{Fig_LL} the Landau levels in graphene corresponding to indices $n=1$ 
and -1 and having the strongest electron-electron interactions, which results in 
a more stable FQHE, are shown by red lines. The electron-electron interaction in the 
Landau level with index $n=0$, 
shown by a blue line, is identical to the interaction in the $n=0$ Landau level of 
the non-relativistic system. Experimental observation of FQHE in a suspended graphene 
\cite{Andrei_2009,Abanin_2010,Kim_2009,Ghahari} and robustness of the FQHE plateaus, 
which were observed even at a weak magnetic field $\sim 2$ Tesla, illustrate the enhancement 
of the electron-electron interactions in graphene as compared to that for a conventional, 
non-relativistic semiconductor systems. 

\section{Interacting Fermions in Bilayer Graphene}

Additional control of the interaction properties of `relativistic' (Dirac-like) 
particles in graphene is also possible in a system of bilayer graphene 
\cite{novoselov_bi,falko_2006,mccann_2006,ohta_2006,koshino_2010} which consists of two 
coupled graphene layers. Bilayer graphene has been intensely investigated because 
of its intriguing properties. The effective low-energy Hamiltonian in this case is
similar to the Dirac-like nature of that in monolayer graphene, but with a quadratic
(instead of linear) off-diagonal term \cite{mccan_chap,book_kats}. The low-energy 
dispersion is also quadratic. The massive Dirac fermions in bilayer graphene also posses 
the pseudospin degree of freedom and are chiral \cite{mccan_chap,book_kats}. In a 
perpendicular magnetic field the Landau levels in a bilayer graphene follow the 
sequence, $\varepsilon^{}_n\propto\sqrt{|n|\left(|n|-1\right)}$ for $n \geq1$ with 
a doubly degenerate $\varepsilon^{}_0=0$ for $n=0$ \cite{falko_2006,zhao_2010}. An 
important characteristic of bilayer graphene is that it is a semiconductor with 
a tunable bandgap between the valence and conduction bands \cite{peeters_07}. This 
property modifies the Landau level spectrum and influences the role of long-range 
Coulomb interactions \cite{abergel}. In a magnetic field, the electronic properties 
of the graphene bilayer can be controlled by (i) the inter-layer bias voltage applied 
to two graphene monolayers, (ii) the intra-layer asymmetry potential due to the
contact of one of the layers with a substrate, (iii) by applying an in-plane magnetic 
field, and (iv) by introducing mechanical deformation of bilayer graphene \cite{Falko_11}. 
Below we consider only the effects of a bias potential and an in-plane magnetic field. 

\section{Biased bilayer graphene}

\begin{figure}
\begin{center}
\includegraphics[width=.5\textwidth]{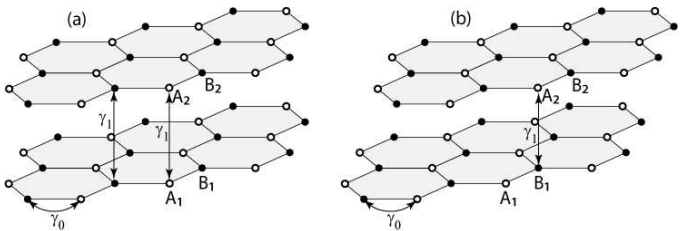}
\end{center}
\caption{\label{bilayer1}
Schematic illustration of two different types of stacking of bilayer graphene, 
consisting of two coupled monolayers of graphene: (a) AA stacking; (b) 
Bernal stacking. Each graphene layer consists of two inequivalent 
sites A and B. The intra-layer and intra-layer hopping 
integrals are shown by $\gamma^{}_0$ and $\gamma^{}_1$, respectively. 
}
\end{figure}

Bilayer graphene comprises two coupled graphene monolayers \cite{falko_2006}. 
Depending on the orientation of the monolayers, there are two main stacking of a 
graphene bilayer: (i) the AA stacking and (ii) the Bernal (AB) stacking, which are shown 
schematically in Fig.~\ref{bilayer1}. There is also the intermediate type of stacking 
of two monolayers corresponding to the rotated bilayer graphene, in which monolayers are
rotated relative to each other by an arbitrary angle \cite{bi_rotate,latil,lopes_07,mele_10}. 
These systems show rich low-energy physics due to the modulated nature of the 
interlayer coupling. 

%-------------------------------------------------------------------- 
%--------------------------------------------------

For the AA stacking [Fig.~2 (a)] in a perpendicular magnetic field, the interlayer 
coupling occurs between the Landau levels of the two layers with the same Landau 
level indices. This coupling changes the energies of the Landau levels of the 
monolayers, but does not affect the wave functions of the layers. Therefore, the 
pseudopotentials, 
which characterize the electron-electron interaction properties, are completely 
identical to the corresponding pseudoptentials of a monolayer graphene. In this 
case the FQHE in the Landau levels of bilayer graphene with AA stacking has the same 
properties, e.g., the same FQHE gaps, as in the corresponding graphene monolayers. 

In the case of Bernal (AB) stacking [Fig.~2 (b)], the interlayer coupling strongly 
modifies the properties 
of the Landau levels in the system. The Hamiltonian of the graphene bilayer with AB 
staking for valley $\xi =\pm 1$ can be written as \cite{mccann_2006} 
\begin{equation}
{\cal H}_{\xi}^{(AB)} = \xi\left( 
\begin{array}{cccc}
\frac{U}2  & v^{}_{\rm F} \pi^{}_{-} & 0 & 0 \\
 v^{}_{\rm F} \pi^{}_{+} & \frac{U}2 &
\xi\gamma^{}_1 & 0 \\ 
 0 &\xi\gamma^{}_1 & -\frac{U}2 &
v^{}_{\rm F}
\pi^{}_{-} \\      
 0 & 0 & v^{}_{\rm F} \pi^{}_{+} & -\frac{U}2 
 \end{array} 
\right),
\label{HAB2}
\end{equation}
where $U$ is the inter-layer bias voltage and $\gamma^{}_1 \approx 400$ meV is the 
interlayer hopping integral. The corresponding wave function is described by a 
four-component spinor $(\psi^{}_{A^{}_1}, \psi^{}_{B^{}_1}, \psi^{}_{B^{}_2}, 
\psi^{}_{Ai^{}_2})^{T}$ for valley $K$ and $(\psi^{}_{B^{}_2}, \psi^{}_{A^{}_2}, 
\psi^{}_{A_1}, \psi^{}_{B^{}_1} )^{T}$ for valley $K^{\prime}$. Here the
sub-indices $A^{}_1$, $B^{}_1$, and $A^{}_2$, $B^{}_2$ correspond to lower and upper 
layers respectively. The wave function corresponding to the Hamiltonian (\ref{HAB2}) 
has the form
\begin{equation}
\Psi^{\rm (bi)}_{n,m}  = 
\left( \begin{array}{c}
 \xi  C^{}_1 \phi^{}_{n-1,m} \\
   C^{}_2   \phi ^{}_{n,m} \\  
   C^{}_3  \phi^{}_{n,m} \\
  \xi C^{}_4 \phi ^{}_{n+1,m}  
\end{array}  
 \right),
\label{fAB2}
\end{equation} 
where $C^{}_1$, $C^{}_2$, $C^{}_3$, and $C^{}_4$ are constants. Therefore, the wave 
functions in bilayer graphene with Bernal stacking is a mixture of the 
non-relativistic Landau wave functions with indices $n-1$, $n$, and $n+1$. 

In the expression for the wave function (\ref{fAB2}) of bilayer graphene, the 
Landau index $n$ can take the following values: $n=-1,0,1,\ldots$. Here we assume 
that if the index of the Landau wave function is negative then the function is 
identically zero, i.e., $\phi^{}_{-2,m} \equiv 0$ and $\phi^{}_{-1,m} \equiv 0$. In 
this case, for $n=-1$ the wave function (\ref{fAB2}) is just $\Psi^{\rm (bi)}_{-1,m} 
= (0,0,0,\phi ^{}_{0,m})$, i.e., the coefficients $C^{}_1$, $C^{}_2$, $C^{}_3$ are 
equal to zero. There is only one energy level corresponding to $n=-1$. For $n=0$, the 
wave function (\ref{fAB2}) has a zero coefficient $C^{}_1$, which results in three 
energy levels corresponding to $n=0$. For other value of $n$, i.e., for $n> 0$, there 
are four eigenvalues of the Hamiltonian (\ref{HAB2}), corresponding to four Landau 
levels in a bilayer for a given valley $\xi = \pm 1$. The eigenvalue equation 
determining these Landau levels, have the form \cite{peeters_07} 
\begin{equation}
\left[\left(\varepsilon+\xi u \right)^2-2n\right]
\left[\left(\varepsilon-\xi u \right)^2-2(n +1)\right] = 
\tilde{\gamma}_1^2 \left[\varepsilon^2-u^2\right],
\label{eigen1}
\end{equation}
where $\varepsilon$ is the energy of the Landau level in units of $\varepsilon^{}_B$ 
($\varepsilon^{}_B=\hbar v^{}_{\rm F}/\ell^{}_0), \ u=U/(2\varepsilon^{}_B)$, and 
$\tilde{\gamma}^{}_1=\gamma^{}_1/ \varepsilon^{}_B$. It is convenient to introduce the 
following labeling scheme for the Landau levels determined by Eq.~(\ref{eigen1}). 
The four Landau levels correspond to two valence levels which have negative energies, 
and two conduction 
levels, which have positive energies. Then the four Landau levels of bilayer graphene 
for a given value of $n$  and a given velley $\xi$ can be labelled as $n^{(\xi)}_i$, 
where $i = -2, -1, 1, 2$ is the label of the Landau level in the ascending order. Here 
negative and positive values of $i$ correspond to the valence and conduction levels, 
respectively. The Landau levels of different valleys are related through the following 
equation $\varepsilon (n^{(\xi)}_i) = - \varepsilon (n^{(-\xi)}_{-i})$. Although for $n=0$ 
there are only three Landau levels and for $n=-1$ there is only one Landau level, 
it is convenient to include the $n=-1$ Landau level into the set of $n=0$ Landau levels 
and label them as $0^{(\xi)}_i$, where $i=-2,-1,1,3$. 

At the zero bias voltage, the Landau levels become two-fold valley and two-fold 
spin degenerate and are given by the expression 
\begin{equation}
\varepsilon = \pm \sqrt{ 2n+1 + \frac{\tilde{\gamma}^{2}_1}{2}\pm 
\frac12 \sqrt{(2+ \tilde{\gamma}^{2}_1 )^2 + 8 n \tilde{\gamma}^{2}_1}}.
\label{zero}
\end{equation}
Since the FQHE is expected only in the Landau levels with low values of the
index, $n$, we consider below the sets of Landau levels of bilayer graphene 
with $n=0$ and $n=1$ only. The wave functions of these Landau levels are 
mixtures of the conventional, non-relativistic Landau functions with indices 
$0$, $1$, and $2$. 

Once the wave functions (\ref{fAB2}) of the bilayer Landau levels are
obtained, the form factor in the pseudopotentials (\ref{Vm}) can be obtained from
\begin{eqnarray}
\nonumber
F^{}_n(q) &=& |C^{}_1|^2 L^{}_{n-1}(q^2/2) +  
\left(|C^{}_2|^2 + |C^{}_3|^2\right) L^{}_{n}(q^2/2)\\ 
&+& |C^{}_4|^2 L^{}_{n+1}(q^2/2).
\label{FFbilayer}
\end{eqnarray}
With the known form factors, the pseudopotentials, which determine the 
interaction strength and the FQHE in a given Landau level, can be calculated.

There are two special Landau levels of bilayer graphene. For $n=-1$ there are  
two solutions (one for the valley $K$ and one for $K^\prime$) of 
Eq.~(\ref{eigen1}) with energies $\varepsilon = -\xi u$. The corresponding 
wave function 
\begin{equation}
\Psi^{\rm (bi)}_{0^{(+)}_1,m}=
\Psi^{\rm (bi)}_{0^{(-)}_{-1},m} =
\left( \begin{array}{c}
0 \\
0 \\
0 \\
\phi^{}_{0,m}
\end{array}
\right),
\label{f00}
\end{equation}
is determined only by the $n=0$ non-relativistic Landau level wave function. 
Therefore the FQHE and the interaction properties of these Landau levels are 
exactly the same as those for the $0$-th conventional (non-relativistic) 
Landau level. 

For $n=0$ and for small values of $u$ there is another solution of 
Eq.~(\ref{eigen1}) with almost zero energy, $\varepsilon\approx 0$. The wave function 
of  this  Landau level has the form 
\begin{eqnarray}
\nonumber
\Psi^{\rm (bi)}_{0^{(+)}_{-1},m} =
\Psi^{\rm (bi)}_{0^{(-)}_{1},m} &=&
\frac1{\sqrt{\tilde{\gamma}_1^2 + 2}}
\left( \begin{array}{c}
0 \\
\sqrt{2} \phi^{}_{0,m} \\
0 \\
\tilde{\gamma}^{}_1 \phi^{}_{1,m}
\end{array}
\right)\\
& = &\frac1{\sqrt{\gamma_1^2 + 2 \epsilon_B^2}}
\left( \begin{array}{c}
0 \\
\sqrt2 \epsilon^{}_B\phi^{}_{0,m} \\
0 \\
\gamma^{}_1 \phi^{}_{1,m}
\end{array}
\right).
\label{f11}
\end{eqnarray}
For a small magnetic field, $\varepsilon^{}_B \ll \gamma^{}_1$, the 
wave function becomes $(\psi^{}_{1,m}, 0, 0, 0)^T$ and the 
Landau level becomes identical to the $n=1$ non-relativistic 
Landau level. In a large magnetic field $\varepsilon^{}_B \gg \gamma^{}_1$, the 
Landau level wave function becomes $(0, 0, \psi^{}_{0,m}, 0)^T$ and the 
bilayer Landau level has the same properties as for the
$n=0$ non-relativistic Landau level. 
The corresponding form factor of the Landau level (\ref{f11}) is given by
\begin{equation}
F^{}_{0^{}_{-1}}(q)=\left[\frac{\gamma_1^2}{\gamma_1^2+2\varepsilon_B^2}\right] 
L^{}_1(q^2/2) + \left[\frac{2\varepsilon_B^2}{\gamma_1^2+2\varepsilon_B^2 }
\right] L^{}_0(q^2/2).
\label{FB01}
\end{equation}
With increasing magnetic field, i.e., with increasing $\varepsilon^{}_B$, 
the bilayer Landau level $0^{}_{-1}$ becomes identical to (i) the
$n=1$ non-relativistic Landau level with the form factor of $L^{}_1(q^2/2)$ 
for small $B$, $\varepsilon^{}_B \ll \gamma^{}_1$; (ii) the $n=1$ 
Landau level of the monolayer graphene with the form factor of $\frac12 
[L^{}_0(q^2/2) + L^{}_1(q^2/2)]$ for $\varepsilon^{}_B=\gamma^{}_1/\sqrt{2}$; 
and, (iii) the $n=0$ non-relativistic Landau level with the form factor of 
$L^{}_0(q^2/2)$ for large $B$, $\varepsilon^{}_B \gg \gamma^{}_1$. 
For typical values of the interlayer coupling, $\gamma^{}_1 = 400$ meV, the condition 
$\varepsilon^{}_B = \gamma^{}_1/\sqrt{2}$ is achieved for a large magnetic field   
$B=120$ Tesla. Under this condition only the first regime can be experimentally realized. 
Below we show that an in-plane magnetic field can suppress the interlayer coupling, 
which opens the possibility of experimental observation of transitions between 
the different regimes (i)-(iii). 

For all the Landau levels [except the levels described by Eq.\ (\ref{f00})]
in bilayer graphene the electron-electron interaction strength and 
the stability, i.e., the excitation gaps of the FQHE states depend on the 
magnetic field $B$, the bias voltage $U$, and the Landau level index. 
Therefore the interaction properties of a bilayer graphene can be controlled by 
the external parameters \cite{bi_FQHE}, which is a different situation 
than in a monolayer graphene, where the interaction properties depend only on 
the Landau level index. 

The stable FQHE states in a bilayer graphene are expected for the $n=0$ and $n=1$ 
Landau level sets, which are the mixtures of the $n=0$, $n=1$, and $n=2$ 
non-relativistic Landau level wave functions. This mixture depends on the 
values of the parameters of the system. To characterize the stability of the 
FQHE we evaluate numerically the FQHE excitation gaps for a finite size system 
in a spherical geometry. We present below the results for the $\nu=\frac13$ FQHE state. 
A similar behavior is expected for other main fractions of the FQHE, i.e., 
$\nu=\frac15$, $\frac25$, $\frac23$ etc.

\begin{figure}
\begin{center}
\includegraphics[width=.5\textwidth]{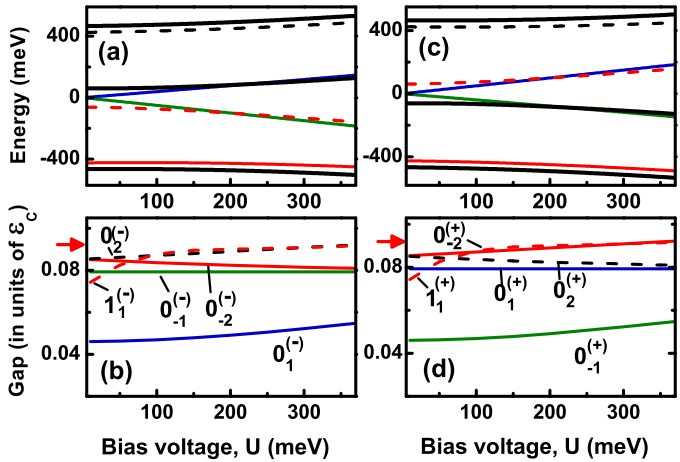}
\end{center}
\caption{\label{FigM1}
The Landau levels of the bilayer graphene [panels (a) and (c)] shown as a 
function of the bias voltage, $U$. Panels (b) and (b): the Coulomb gaps of 
the $\frac13-$FQHE in corresponding Landau levels. The results are obtained 
in spherical geometry for a finite-size system with eight electrons and
$2S=21$ flux quanta. The numbers next to the lines 
are the labels of the Landau levels. The same type of lines [in panels 
(a) and (b) and panels (c) and (d)] correspond to the same Landau levels.
Panels (a) and (b) correspond to the valley $K^{\prime}$, while panels (c) and (d) 
correspond to the valley $K$. The system is characterized by 
$\gamma^{}_1 = 400$ meV and a magnetic field $B=15$ 
Tesla.  The arrows in panels (b) and (d) indicate the 
gap of the $\frac13-$FQHE in the $n=1$ Landau level of a monolayer graphene.
}
\end{figure}

In Fig.\ \ref{FigM1} we show the dependence of the Landau levels on the bias voltage 
$U$ for a fixed magnetic field and for different valleys. The results are presented only 
for the Landau levels with indices $n=0$ and 1, i.e. only for the Landau level where 
the FQHE can be observed. The corresponding $\nu = \frac13$ FQHE gaps are shown in 
Fig.\ \ref{FigM1} (b,d). Both for the $K$ and $K^{\prime }$ valleys there is a special 
Landau level, $O_{1}^{(+)}$ (for the $K$ valley) and $O_{-1}^{(-)}$ (for the $K^{\prime}$ 
valley), that is described by the wave function of the type as in Eq.\ (\ref{f00}). In 
these Landau levels the FQHE gap does not depend on the bias voltage and is exactly 
equal to the FQHE gap of a conventional (non-relativistic) semiconductor systems for the 
$n=0$ Landau level. In all other levels the FQHE gap depends on the bias voltage, which 
clearly illustrates the sensitivity of the interaction properties on the external 
parameters, i.e., the bias voltage. Although the interaction strength within a single 
Landau level can be controlled by the bias voltage, the results illustrated in 
Fig.\ \ref{FigM1} show that the FQHE gaps in bilayer graphene are usually less than 
the largest FQHE gap in a monolayer graphene. This FQHE gap in a monolayer graphene 
is realized in the $n=1$ Landau level and is shown by red arrows in Fig.\ \ref{FigM1}. 

%-------------------------------------------------------------------- 
%-----------------------------------------------------------

\begin{figure}
\begin{center}
\includegraphics[width=.5\textwidth]{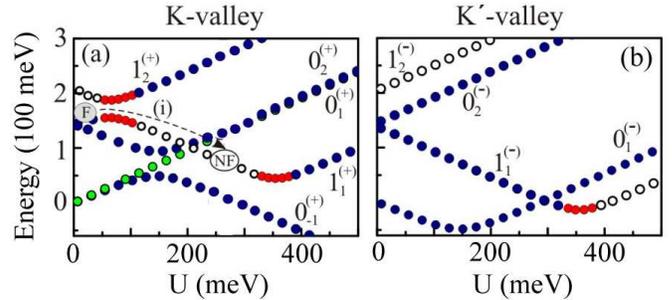}
\end{center}
%\centering
%\includegraphics[width=.8\textwidth]{..//Figures/bilayer_Fig1_new.eps}
\caption{\label{FigM4}
A few lowest Landau levels of the conduction band as a function of 
the bias potential, $U$, for inter-layer coupling of  $\gamma^{}_1=30$ meV 
and a magnetic field of 15 Tesla. The numbers next to the curves denote 
the corresponding Landau levels. Panels (a) and (b) 
correspond to the $K$ and $K^{\prime}$ valleys, respectively. The 
Landau levels where the FQHE can be observed are drawn as blue and green filled 
dots. The green dots correspond to the Landau levels where the FQHE states are 
identical to that of a monolayer of graphene or a non-relativistic conventional 
system. The red dots represent Landau levels with a weak FQHE. The open dots correspond 
to Landau levels where the FQHE is absent. In (a), the dashed lines labeled by 
(i) illustrates the transition between FQHE (symbol `F') and no FQHE (symbol `NF') 
states  under a constant gate voltage and variable bias potential \cite{bi_FQHE}.
}
\end{figure}

For a smaller interlayer hopping integral, the Landau levels in bilayer graphene show 
anticrossings as a function of the bias voltage \cite{bi_FQHE}. These anticrossings 
result in a strong mixture of different Landau levels, which can greatly modify the 
properties of the Landau level wave functions and change the interaction strength 
within a single Landau level. This behavior is illustrated in Fig.\ \ref{FigM4}, 
where the dependence of the Landau levels on the bias voltage is shown for 
$\gamma^{}_1 = 30$ meV. The anticrossings of the Landau levels result in transitions 
from an incompressible state (FQHE) to a compressible state (no FQHE) within 

a single Landau level (see the Landau level $1^{(+)}_2$ in Fig.\ \ref{FigM4} (a)). 
There is also a double transition, marked by the dashed line (i), at the Landau level 
$1_1^{(+)}$. At this Landau level, the electron system with fractional filling 
shows transitions from an incompressible state (FQHE) at small bias voltage $U$ 
to a compressible state (no FQHE) at intermediate values of $U$ and then to an
incompressible state (FQHE) at large $U$. No such transition has ever occured in 
conventional semiconductor systems.

Although for experimentally realized bilayer systems the interlayer hopping 
integral is relatively large, $\gamma^{}_1 \approx 400$ meV, the interlayer 
coupling can be controlled and suppressed by an applied in-plane magnetic field. 
This situation is discussed in the next section. 

%-------------------------------------------------------------------- 
%-----------------------------------------------------------

\section{Bilayer graphene in a tilted magnetic field}

A tilted magnetic field, applied to a quasi-two-dimensional electron system, 
can modify the electron dynamics and correspondingly the electron-electron interaction 
strength. In a graphene monolayer, due to its purely 2D nature, the component of the 
magnetic field parallel to the monolayer does not influence the electron's spatial 
dynamics, although it can alter the electron spin dynamics, which is sensitive to the 
total magnetic field \cite{FQHE_spin_theory,FQHE_spin_expt}. 

Bilayer graphene is a quasi-two-dimensional system. The electron dynamics 
in such a system is sensitive to both perpendicular and in-plane components of the 
magnetic field \cite{tilted1}. To introduce a tilted magnetic field into the Hamiltonian 
of bilayer graphene, we introduce the vector potential 
$\vec{A} = (0,B^{}_{\perp}x,B^{}_{\parallel}y)$, where $B^{}_{\perp}$ ($z$-component) and 
$B^{}_{\parallel}$ ($x$-component) are perpendicular and in-plane components of the 
tilted magnetic field. Here the $z$ axis is perpendicular to the graphene monolayers. 
The perpendicular component of the magnetic field, which alters the electron dynamics 
in the $(x,y)$ plane, is introduced in the bilayer Hamiltonian by replacing the 2D 
momentum by the generalized momentum [similar to the Hamitonian (\ref{HAB2})]. The parallel 
component of the magnetic field is introduced through the Peierls substitution by 
multiplying the interlayer hopping integral $\gamma^{}_1$ with the phase factor 
$\exp (-{\rm i}e/\hbar A^{}_z d )=\exp(-{\rm i}\beta y)$, where $\beta = eB^{}_{\parallel} 
d/\hbar$, and $d$ is the interlayer separation. Then the Hamiltonian of the bilayer graphene 
with AB stacking and at zero bias voltage becomes
\begin{equation}
{\cal H}_{\xi}^{(AB)} = \xi\left( 
\begin{array}{cccc}
0  & v^{}_{\rm F} \pi^{}_{-} & 0 & 0 \\
 v^{}_{\rm F} \pi^{}_{+} & 0 &
\xi\gamma^{}_1 {\rm e}^{- i \beta y} & 0 \\ 
 0 &\xi\gamma^{}_1  {\rm e}^{ i \beta y}& 0 &
v^{}_{\rm F}
\pi^{}_{-} \\      
 0 & 0 & v^{}_{\rm F} \pi^{}_{+} & 0 
 \end{array} 
\right).
\label{Htilted}
\end{equation}
The wave functions corresponding to the Hamiltonian (\ref{Htilted}) can be expressed 
in terms of the non-relativistic 2D Landau wave functions. For the vector potential 
$\vec{A}^{}_{\perp} = (0,B^{}_{\perp}x,0)$ corresponding to the perpendicular component of 
the magnetic field, the 2D Landau wave functions are parametrized by the $y$ component of 
the wave vector and the Landau index $n$, and are described as
\begin{eqnarray}
\nonumber
 \phi^{}_{|n|,k} (x,y) \propto {\rm e}^{{\rm i}ky} \psi^{}_k (x) 
&=& C^{}_0 {\rm e}^{{\rm i}ky} H^{}_{|n|}
 \left( \frac{x-x^{}_k}{\ell^{}_0} \right)\\
&& \times\exp \left[- \frac{(x-x^{}_k)^2}{2\ell_0^2}\right], 
\label{psiLL}
\end{eqnarray}
where $\psi^{}_k(x) = C^{}_0  H^{}_{|n|}\left( \frac{x-x^{}_k}{\ell^{}_0} \right)
\exp \left[- \frac{(x-x^{}_k)^2}{2\ell _0^2}\right]$, $H^{}_n(x)$ are the Hermite polynomials, 
and $x^{}_k = k\ell_0^2$. Here the magnetic length $\ell^{}_0$ is defined by the perpendicular 
component of the magnetic field, $\ell^{}_0 = \sqrt{\hbar / eB^{}_{\perp}}$.
Then the wave functions of the Hamiltonian (\ref{Htilted}) are parametrized 
by the Landau level index $n$ and the wave vector $k$ and have the form
\begin{equation}
\Psi^{\rm (bi)}_{n,k}  = 
\left( \begin{array}{c}
 \xi  C^{}_1 \phi^{}_{|n|-1,k} \\
   C^{}_2   \phi ^{}_{|n|,k-\beta} \\  
   C^{}_3  \phi^{}_{|n|,k+\beta} \\
  \xi C^{}_4 \phi ^{}_{|n|+1,k}  
\end{array}  
 \right),
\label{ftilted}
\end{equation} 
where $C^{}_1$, $C^{}_2$, $C^{}_3$, and $C^{}_4$ are constants. The in-plane component 
of the magnetic field results in a coupling of the Landau wave functions with 
the wave vectors $k$, $k-\beta$, and $k+\beta$. For the wave functions of the form 
(\ref{ftilted}) the Hamiltonian of bilayer graphene in a Landau level 
with index $n$ takes the form  
\begin{equation}
{\cal H}_{\xi,n}^{(AB)} = \xi\left( 
\begin{array}{cccc}
0  & v^{}_{\rm F} \pi^{}_{-} & 0 & 0 \\
 v^{}_{\rm F} \pi^{}_{+} & 0 &
\xi\gamma^{}_1 \kappa^{}_n(\mu ) & 0 \\ 
 0 &\xi\gamma^{}_1  \kappa^{}_n(\mu ) & 0 &
v^{}_{\rm F}
\pi^{}_{-} \\      
 0 & 0 & v^{}_{\rm F} \pi^{}_{+} & 0 
 \end{array} 
\right),
\label{Htilted2}
\end{equation}
where 
\begin{equation}
\kappa^{}_n (\mu )= \int dx \psi^{}_k (x) \psi^{}_{k-\beta} (x) 
\end{equation}
depends on the dimensionless parameter $\mu = \beta \ell^{}_0 = eB^{}_{\parallel} d \ell^{}_0 
/\hbar$. Therefore the effect of the in-plane component of the magnetic field on the 
electron dynamics in a bilayer graphene is the reduction of the interlayer coupling, 
i.e., the interlayer coupling is $\gamma^{}_{1,n} = \gamma^{}_1 \kappa^{}_n(\mu)< \gamma^{}_1$. 
This reduction depends on the Landau level index and on the dimensionless parameter $\mu $. 
Due to the small interlayer distance $d = 3.3$ \AA, the parameter $\mu $ is relatively 
small. To increase the value of this parameter, the perpendicular component 
of the magnetic field needs to be small, i.e., the magnetic length should be large, and the 
parallel component of the magnetic field should also be large. For example, for $B^{}_{\perp} 
= 1$ Tesla, $\mu = 0.014 B^{}_{\parallel} {[\rm Tesla]}$. For the first lowest Landau level 
indices the function $\kappa^{}_n(\mu)$ is $\kappa^{}_0 (\mu) = {\rm e}^{-\mu^2/4}$ 
and $\kappa^{}_1 (\mu) = {\rm e}^{-\mu^2/4} \left( 1- \frac{\mu^2}{2}\right)$. 

%--------------------------------------------------------------------
\begin{figure}
\begin{center}
\includegraphics[width=.5\textwidth]{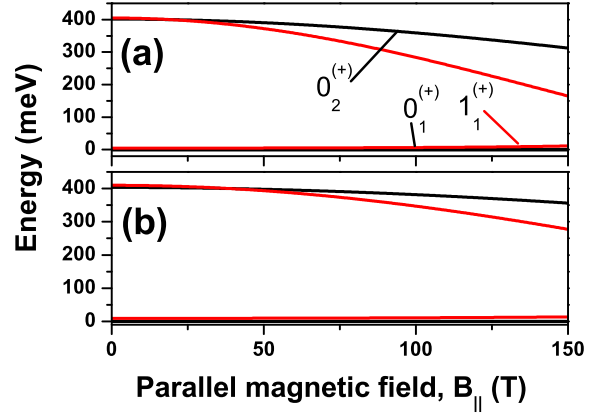}
\end{center}
%\centering
%\includegraphics[width=.4\textwidth]{..//Figures/bilayer_Fig3_new.eps}
\caption{\label{Fig_tilted1}
The Landau levels of bilayer graphene in a tilted magnetic field and zero bias
voltage shown as a function of the parallel component of the magnetic field. The
perpendicular component of the tilted magnetic field is (a) 1 Tesla and (b) 2 Tesla.
The labels next to the lines denote the corresponding Landau levels, where only
the Landau levels in which the FQHE can be observed, are labelled. Only the Landau
levels with positive energies are shown.
}
\end{figure}
%-----------------------------------------------------------

The Landau levels of a bilayer graphene in a tilted magnetic field are given by 
Eq.~(\ref{zero}), in which the interlayer coupling $\tilde{\gamma}^{}_1$ should be replaced 
by $\tilde{\gamma}^{}_{1,n} = \gamma^{}_{1,n}/\varepsilon^{}_B$. Here $\varepsilon^{}_B$ is 
calculated in terms of the perpendicular component of the tilted magnetic field. In 
Fig.\ \ref{Fig_tilted1} the dependence of the Landau levels on the parallel component 
of the magnetic field, $B^{}_{\parallel}$, is shown for a few lowest Landau levels of a bilayer 
graphene. Increasing the parallel component of the magnetic field, the energies of 
the Landau levels are reduced, which is consistent with the reduction of the interlayer 
coupling, $\gamma^{}_{1,n}$, with increasing $B^{}_{\parallel}$. The dependence of the Landau 
levels on $B^{}_{\parallel}$ becomes weaker with increasing perpendicular magnetic field 
[Fig.\ \ref{Fig_tilted1} (a,b)]. Therefore, the effect of an in-plane magnetic 
field on the Landau levels can be observed only for a small perpendicular magnetic field, 
$B^{}_{\perp} \approx 1 $, and a large parallel magnetic field, $B^{}_{\parallel}\geq 50$
Tesla. It should be pointed out that although the perpendicular component of the field is 
rather small, in a conventional semiconductor system the FQHE has been reported in a magnetic 
field of $B < 3$ Tesla \cite{igor}.

\begin{figure}
\begin{center}
\includegraphics[width=8cm]{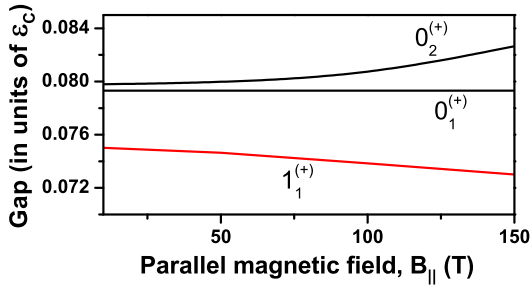}
\end{center}
%\centering
%\includegraphics[width=.8\textwidth]{..//Figures/bilayer_Fig3_new.eps}
\caption{\label{Fig_tilted2}
The FQHE gaps at different Landau levels of bilayer graphene shown as a 
function of the parallel component of the tilted magnetic field. The 
perpendicular component of the magnetic field is 1 Tesla. The labels next 
to the lines denote the corresponding Landau levels [Fig.\ \ref{Fig_tilted1}]. 
The bias voltage is zero in this case. The results are obtained in spherical
geometry for a finite-size system with eight electrons and $2S=21$ 
flux quanta.
}
\end{figure}

The interaction properties of electrons in the Landau levels of a bilayer graphene 
also depend on the in-plane component of the magnetic field. This dependence 
is visible only for small perpendicular components of the magnetic field, i.e., 
$B^{}_{\perp} \approx 1 $ Tesla. In Fig.\ \ref{Fig_tilted2} we show the $\frac13$-FQHE gap 
as a function of the in-plane component of the magnetic field for different Landau 
levels. For $B^{}_{\perp}\approx 1$ Tesla, only three Landau levels (with positive energies) 
support the FQHE states. One Landau level $0^{(+)}_1$, the wave function of which has 
the form of (\ref{f00}) and depends only on the perpendicular component of the magnetic 
field, does not show any dependence on the in-plane component of the magnetic field. 
The interaction strength in the Landau levels $0^{(+)}_2$ and $1^{(+)}_1$ depends weakly 
on $B^{}_{\parallel}$ [Fig.\ \ref{Fig_tilted2}]. The interaction strength 
increases with $B^{}_{\parallel}$ for the Landau level $0^{(+)}_2$ and decreases for the 
Landau level $1^{(+)}_1$. Therefore the parallel component of the magnetic field 
can in fact, {\it enhance} the electron-electron interaction strength for some Landau 
levels (Fig.\ \ref{Fig_tilted2}) in a bilayer graphene. 

%-------------------------------------------------------------------- 
%-----------------------------------------------------------

\section{The pfaffians in graphene}

For conventional 2D semiconductor systems, in addition to the usual incompressible 
FQHE states that are realized for the odd-denominator filling factors, the 
electron-electron interaction is also responsible for the formation of a special 
type of incompressible state corresponding to the fractional filling factor 
$\nu = \frac52$. This filling factor corresponds to a completely occupied $n=0$ 
Landau level with two spin components and a half-filled $n=1$ Landau level. Since 
the completely occupied Landau levels do not contribute to the electron dynamics,
the ground state of the $\nu = \frac52$ system is determined by the electron-electron 
interaction alone in the $n= 1$ Landau level. The incompressible state with a
large excitation gap is formed in this half-filled Landau level. One unusual property
of this state is that the elementary charged excitations have a charge $e^*=e/4$. 
They obey the ``non-abelian" statistics \cite{halperin,stern} and carry the signature 
of Majorana fermions \cite{majorana}. It was proposed that the ground state of the
half-filled $n=1$ Landau level is described by a Pfaffian \cite{read,greiter} 
or the anti-Pfaffian function \cite{anti1,anti2}. The Pfaffian state is written
\begin{equation}
\Psi^{}_{\mbox{Pf}}=\mbox{Pf}\left(\frac1{z^{}_i-z^{}_j}\right)
\prod_{i<j} (z^{}_i-z^{}_j)^2 \exp\left(-\sum_i\frac{z_i^2}{4\ell_0^2}\right),
\end{equation}
where the positions of the electrons are described in terms of the complex variable 
$z= x-{\rm i}y$ and the Pfaffian is defined as \cite{read,greiter}
\begin{equation}
{\rm Pf}\,M^{}_{ij}=\frac1{2^{N/2}\left(N/2\right)!}\sum_{\sigma
\in S^{}_N}{\rm sgn}\,\sigma\prod_{l=1}^{N/2}M^{}_{\sigma(2l-1)\sigma(2l)},
\end{equation}
for an $N\times N$ antisymmetric matrix whose elements are $M^{}_{ij}$. Here
$S^{}_N$ is the group of permutations of $N$ objects.

The Pfaffian state is the exact ground state with zero energy for the electron 
system at half filling with a special {\em three-particle} interaction which is 
non-zero only if all three particles are in close proximity to each other 
\cite{greiter_2}. For realistic two-particle interactions the Pfaffian state is 
not an exact eigenstate of the half-filled system. In the case of the Coulomb interaction, 
the overlap of the ground state of the $\nu = \frac12$ system in the $n=1$ Landau 
level with the Pfaffian function is around 80\%. By varying the two-particle interaction 
potential, i.e., the pseudopotentials, a stronger overlap ($\sim 99$\%) of the 
ground state of the $\nu = \frac12$ system in the $n=1$ Landau level with the Pfaffian 
state is possible. The proximity of the actual $\nu=\frac12$ ground state to 
the Pfaffian state is most sensitive to the lowest pseudopotentials, $V^{}_1$, 
$V^{}_3$, and $V^{}_5$. In graphene, there are two lowest Landau levels with indices 
$n= 0$ and $n=1$ with strong electron-electron interactions. Although conventional 
FQHE states with odd-denominator filling factors can be observed at these Landau 
levels, the Pfaffian state with half-filling of the corresponding Landau level 
cannot be realized \cite{Pfaffian_PRL}. In the $n=0$ Landau level in graphene, the 
interaction potential is identical to the one in the $n=0$ non-relativistic Landau
level, and similar to the case of the non-relativistic system, the Pfaffian state 
is not the ground state of the $\nu = \frac12$ system in the $n=0$ Landau level 
in graphene. In the $n= 1$ Landau level in graphene, although the electron-electron 
interaction results in the stable odd denominator FQHE states, the ground state 
of the half-filled Landau level is compressible and is not described by the Pfaffian 
function. The overlap of the ground state of the $\nu= \frac12$ system with the 
Pfaffian function is less than 0.5 in all Landau levels of the monolayer graphene
\cite{Pfaffian_PRL}. In bilayer graphene the interaction strength and the corresponding 
Haldane pseudopotentials can be controlled by the external parameters, such as the
bias voltage and the direction of the magnetic field. In this case the stability (i.e., 
the magnitude of the excitation gap) of the Pfaffian state can be strongly enhanced.
In bilayer graphene there are two 'special' Landau levels $0_{-1}^{(+)}$ (for valley 
$K$) and $0_1^{(-)}$ (for valley $K^{\prime }$), that are described by Eq.~(\ref{f11}). 
The numerical calculations in a spherical geometry show that
only at these special Landau levels the overlap of the ground state with the Pfaffian 
state is large \cite{Pfaffian_PRL}. In all other bilayer Landau levels the overlap of the 
$\nu=\frac12$ ground state with the Pfaffian state is found to be small ($< 0.6$) and 
these states cannot be described by the Pfaffian function. In spherical geometry, 
the Pfaffian state in a system of $N^{}_e$ electrons is realized for the parameter 
$2S=2N^{}_e-3$, which corresponds to filling factor $\nu=\frac12$ in the thermodynamic 
limit.

For the zero bias voltage, the energies of the Landau levels $0_{-1}^{(+)}$ and $0_1^{(-)}$
[Eq.~(\ref{f11})] are zero and the levels are degenerate with zero-energy Landau level
given by Eq.~(\ref{f00}). For a finite bias voltage, the degeneracy of the Landau levels 
is lifted. The form factor $F^{}_n$, in the Landau level $0_{-1}^{(+)}$ is calculated 
from Eq.~(\ref{FB01}) and determines the interaction properties of the electron system 
in the level $0_{-1}^{(+)}$. For a small magnetic field, $\gamma ^{}_1 \gg\epsilon^{}_B$, 
the form factor is identical to the form factor of the non-relativistic $n=1$ Landau level. 
Therefore, in this limit the ground state of the $\nu=\frac12$ half-filled system in 
the $0_{-1}^{(+)}$ Landau level is incompressible and is determined by the Pfaffian state.
In a large magnetic field, $\gamma ^{}_1 \ll \epsilon^{}_B$, the form factor $F^{}_n$ 
becomes identical to that of the $n=0$ non-relativistic system, for which the $\nu = 
\frac12$ state is compressible. For intermediate values of the magnetic field, the 
$\nu = \frac12$ system in the $0_{-1}^{(+)}$ Landau level shows an unique behavior as a 
function of the magnetic field: with increasing magnetic field the overlap of the ground 
state of the system with the Pfaffian state shows a maximum for a finite value of the
magnetic field \cite{Pfaffian_PRL}. Therefore, the stability of the Pfaffian $\nu =\frac12$ 
state in bilayer graphene can be increased when compared to that in non-relativistic 
systems.

%--------------------------------------------------------------------
\begin{figure}
\begin{center}\includegraphics[width=7cm]{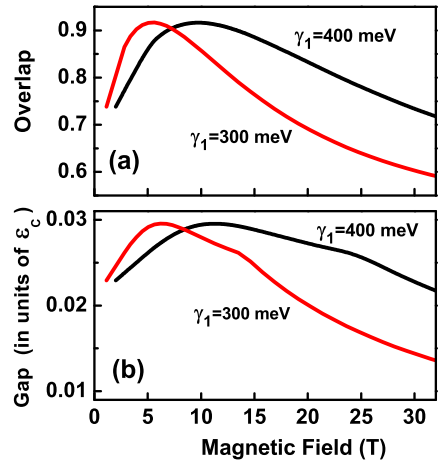}
\end{center}
\caption{\label{FigPfgap1}
(a) Overlap of the exact many-particle ground state with the Pfaffian function. 
(b) Collective excitation gap of the $\nu=\frac12$ state. The results are for 
$N^{}_e=14$, $2S=25$, and $U=5$ meV. The black and red lines correspond to
$\gamma^{}_1 =400$ meV and 300 meV, respectively. The results are shown for the 
$\nu=\frac12$ system in the Landau level $0_{-1}^{(+)}$.
}
\end{figure}
%-----------------------------------------------------------

Our results shown in Fig.~\ref{FigPfgap1} illustrate the non-monotonic dependence 
of the interaction properties of the $\nu=\frac12$ system in the $0_{-1}^{(+)}$ 
Landau level. Here the overlap of the ground state with the Pfaffian state and 
the corresponding excitation gap are shown. With increasing magnetic field the 
properties of the system change non-monotonically and for $\gamma^{}_1 = 400$ meV
the overlap with the Pfaffian state has a maximum in a magnetic field of $\sim10$ 
Tesla. The corresponding excitation gap also has a maximum at this point.
In dimensionless units the maximum appears when $\gamma^{}_1/\epsilon^{}_B 
\approx 4.9$. Therefore, for smaller values of $\gamma^{}_1$, the maximum of the 
overlap is realized for a smaller value of magnetic fields (see the results for 
$\gamma^{}_1 = 300$ meV in Fig.~\ref{FigPfgap1}).

%--------------------------------------------------------------------
\begin{figure}
\begin{center}\includegraphics[width=7cm]{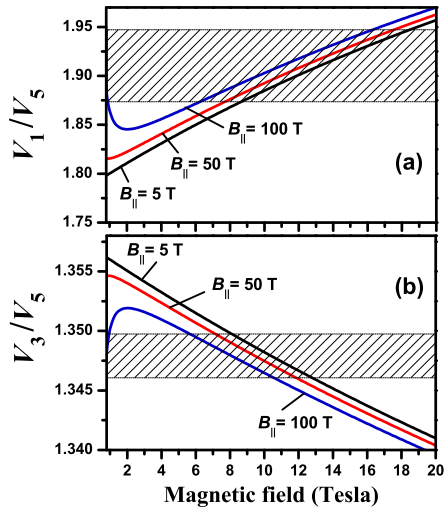}
\end{center}
\caption{\label{FigRatio}
Ratios of pseudopotentials for two values of the angular momentum $V^{}_1/V^{}_5$ 
[panel (a)] and $V^{}_3/V^{}_5$ [panel (b)] as a function of the 
perpendicular component of the magnetic field and for different parallel
components of the magnetic field, $B^{}_{\parallel} = 5$, 50, and 100 Tesla. The 
data are shown for the Landau level $0_{-1}^{(+)}$. The hatched regions 
correspond to the values of the pseudopotentials where one observs a 
large overlap of the ground state with the Pfaffian state and also a large 
excitation gap of the incompressible ground state.
}
\end{figure}
%-----------------------------------------------------------

It is possible to suppress the interlayer hopping integral $\gamma^{}_1$ by applying 
a tilted magnetic field where the in-plane component of the magnetic field determines 
the suppression of $\gamma^{}_1$ by a factor of $\kappa^{}_n$. To identify the effect 
of the in-plane magnetic field on the stability of the Pfaffian state, we characterize 
the interaction properties of the half-filled Landau level by the ratios of the 
pseudopotentials corresponding to the lowest relative angular momentum. The $\nu=\frac12$ 
Pfaffian state is most sensitive to two parameters of the pseudopotentials $V^{}_1/V^{}_5$ 
and $V^{}_3/V^{}_5$ \cite{storni}. In Ref.\ \cite{storni}, in the thermodynamics limit,
the region of the pseudopotenial parameters, for which the most stable Pfaffian state 
can be realized, was obtained in the plane $(V^{}_1/V^{}_5)-(V^{}_3/V^{}_5)$. We apply 
that approach on our bilayer system and evaluate the pseudopotential parameters
$V^{}_1/V^{}_5$ and $V^{}_3/V^{}_5$ in the Landau level $0_{-1}^{(+)}$ as a function 
of the magnetic field. In this way, we can identify the regions of the magnetic field 
with the most stable Pfaffian state. In Fig.\ \ref{FigRatio} the parameters $V^{}_1/V^{}_5$ 
and $V^{}_3/V^{}_5$ are shown as a function of the perpendicular component of the 
magnetic field and for different in-plane components of the magnetic field. These 
results demonstrate that with increasing parallel component of the magnetic field the 
values of the pseudopotentials, which correspond to the most stable Pfaffian state 
and which are illustrated by the hatched region in Fig.\ \ref{FigRatio}, are obtained 
for smaller values of the parallel magnetic field. Therefore, for a given value of 
the perpendicular component of the magnetic field the parallel component of the magnetic 
field increases the stability of the Pfaffian state.

Another interesting effect, introduced by the parallel component of the magnetic field, 
is the strong modification of the interaction properties of the electron system at small 
values of the perpendicular magnetic field. Such changes in the interaction potential 
result in an enhancement of the stability of the Pfaffian states for small values of 
$B^{}_{\perp}$. As an example, for  $B^{}_{\parallel } = 100$ Tesla the stability 
of the Pfaffian state is strongly increased for a weak perpendicular magnetic field, 
$B^{}_{\perp} \lesssim 2$ Tesla (Fig.\ \ref{FigRatio}).

\section{Concluding remarks}

In a magnetic field the strength of the electron-electron interaction, which is 
characterized by the value of the FQHE gap in a given Landau level, depends on 
the Landau level index and the external parameters of the graphene system. In the
case of monolayer graphene, there are two types of Landau levels with indices 
$n=0$ and $n=1$, which have strong electron-electron interactions, i.e., the FQHE 
can be observed only in these Landau levels. Among these Landau levels, the strongest 
electron-electron interactions are realized in the $n=1$ graphene Landau level. In 
the Landau level with index $n=0$, the interaction strength is exactly the 
same as that in the $n=0$ Landau level of the conventional (non-relativistic) 
system, which results in exactly the same FQHE gaps. 

The strength of the electron-electron interaction can be further controlled in a bilayer 
graphene, where the additional parameters that govern the interaction strength 
are the interlayer coupling, the bias voltage, and the orientation of the magnetic 
field. The bias voltage between the graphene monolayers changes the structure of the 
wave functions of the Landau levels and can strongly modify the interaction strength. 
In some Landau levels the electron-electron interaction can be stronger than 
that in a monolayer graphene, resulting in a more stable FQHE. In a given Landau 
level and as a function of the bias voltage the bilayer graphene system can show 
transitions from a state with weak electron-electron interaction (FQHE being absent) 
to a state with strong electron-electron interactions (presence of FQHE). In a bilayer 
graphene, the electron-electron interactions can be additionally controlled by the 
direction of the magnetic field, i.e., in a tilted magnetic field. The sensitivity 
of the interaction strength to the parallel component of the magnetic field is 
visible only for a weak perpendicular component of the magnetic field, $B^{}_{\perp} 
\approx 1$ Tesla and for a strong parallel component of the magnetic field, 
$B^{}_{\parallel} \geq 50$ Tesla. Finally, we describe the stability of the Pfaffian 
state and the excitation gap in a half-filled $n=1$ Landau level in bilayer graphene.
We also discuss the possibility of making the Pfaffian state more stable by applying a 
tilted magnetic field.

\section{Acknowledgments}

This work has been supported by the Canada Research Chairs program of the
Government of Canada.


\begin{thebibliography}{9999}

\bibitem[\ddag]{byline} Electronic address:
tapash@physics.umanitoba.ca

\bibitem{abergeletal}
D.S.L. Abergel, V. Apalkov, J. Berashevich, K. Ziegler, and
T. Chakraborty, Adv. Phys. {\bf 59}, 261 (2010).

\bibitem{ando_07}
T. Ando, Physica E {\bf 40}, 213 (2007).

\bibitem{book_kats}
M.I. Katsnelson, {\it Graphene, Crabon in Two Dimensions}, 
(Cambridge University Press, Cambridge, 2012).

\bibitem{book_raza}
H. Raza (Ed.), {\it Graphene Nanoelectronics} (Springer,
Heidelberg, 2012).

\bibitem{wallace}
P.R. Wallace, Phys. Rev. {\bf 71}, 622 (1947).

\bibitem{geim_rise}
A.K. Geim and K.S. Novoselov, Nat. Mat. {\bf 6}, 183 (2007).

\bibitem{netoetal}
A.H. Castro Neto, F. Guinea, N.M.R. Peres, K.S. Novoselov, and
A.K. Geim, Rev. Mod. Phys. {\bf 81}, 109 (2009).

\bibitem{regan}
M. Mecklenburg and B.C. Regan, Phys. Rev. Lett. {\bf 106}, 116803
(2011); M. Trushin and J. Schliemann, {\it ibid.} {\bf 107},
156801 (2011).

\bibitem{mcclure}
J.W. McClure, Phys. Rev. {\bf 104}, 666 (1956).

\bibitem{haering}
R.R. Haering and P.R. Wallace, J. Phys. Chem. Solids {\bf 3}, 253
(1957).

\bibitem{Novoselov_2005}
K.S. Novoselov, A.K. Geim, S.V. Morozov, D. Jiang, M.I. Katsnelson, 
I.V. Grigorieva, S.V. Dubonos, and A.A. Firsov, Nature {\bf 438},
197 (2005).

\bibitem{Zhang_2005}
Y. Zhang, Y.-W. Tan, H.L. Stormer, and P. Kim,
Nature {\bf 438}, 201 (2005).

\bibitem{haldane1} F.D.M. Haldane, Phys. Rev. Lett. {\bf 51}, 605 (1983).
\bibitem{haldane2} F.D.M. Haldane and E.H. Rezayi, Phys. Rev. Lett. 
{\bf 54}, 237 (1985).

\bibitem{haldane_87}
F.D.M. Haldane, in {\it The Quantum Hall Effect}, Eds. R.E. Prange
and S.M. Girvin (Springer, New York, 1987), p. 303.

\bibitem{Apalkov_2006}
V.M. Apalkov, and T. Chakraborty, Phys. Rev. Lett.
{\bf 97}, 126801 (2006).

\bibitem{Goerbig_06}
M.O. Goerbig, R. Moessner, and B. Ducot, Phys. Rev. B {\bf 74}, 161407(R)
(2006).

\bibitem{FQHE_book}
T. Chakraborty and P. Pietil\"ainen, {\it The Quantum Hall
Effects}, 2nd edition (Springer, New York, 1995).

\bibitem{greiter}
M. Greiter, Phys. Rev. B {\bf 83}, 115129 (2011).

\bibitem{fano} G. Fano, F. Ortolani, and E. Colombo, Phys. Rev. B 
{\bf 34}, 2670 (1986).

\bibitem{Apalkov_2007} V. Apalkov, X.F. Wang, and T. Chakraborty, 
Int. J. Mod. Phys. B {\bf 21}, 1165 (2007).

\bibitem{Toke_2006} C. Toke, P.E. Lammert, V.H. Crespi,
J.K. Jain, Phys. Rev. B 74 (2006), p. 235417.

\bibitem{Shibata_2009} N. Shibata and K. Nomura, J. of Phys. Soc. of Japan
78, 104708 (2009). 

\bibitem{Andrei_2009}
X. Du, I. Skachko, F. Duerr, A. Luican, and E.Y. Andrei,
Nature {\bf 462}, 192 (2009).

\bibitem{Abanin_2010}
D.A. Abanin, I. Skachko, X. Du, E.Y. Andrei, and L.S. Levitov,
Phys. Rev. B {\bf 81}, 115410 (2010).

\bibitem{Kim_2009}
K.I. Bolotin, F. Ghahari, M.D. Shulman, H.L. St\"ormer, and
P. Kim, Nature {\bf 462}, 196 (2009).

\bibitem{Ghahari}
F. Ghahari, Y. Zhao, P. Cadden-Zimansky, K. Bolotin, and P. Kim,
Phys. Rev. Lett. {\bf 106}, 046801 (2011).

\bibitem{novoselov_bi}
K.S. Novoselov, E. McCann, S.V. Morozov, V.I. Fal'ko, M.I. Katsnelson, U. Zeitler, 
D. Jiang, F. Schedin, and  A.K. Geim, Nat. Phys. {\bf 2}, 177 (2006).

\bibitem{falko_2006}
E. McCann and V. Falko, Phys. Rev. Lett. {\bf 96}, 086805 (2006).

\bibitem{mccann_2006} E. McCann, Phys. Rev. B {\bf 74}, 161403 (2006).

\bibitem{ohta_2006}
T. Ohta, A. Bostwick, T. Seyller, K. Horn, E. Rotenberg,
Science {\bf 313}, 951 (2006).

\bibitem{koshino_2010} M. Koshino and E. McCann, Phys. Rev. B {\bf 81},
115315 (2010).

\bibitem{mccan_chap}
E. McCann, in \protect\cite{book_raza}, Ch. 8.

\bibitem{zhao_2010}
Y. Zhao, P. Cadden-Zimansky, Z. Jiang, and P. Kim, Phys. Rev. Lett.
{\bf 104}, 066801 (2010).

\bibitem{peeters_07}
J.M. Pereira, Jr., F.M. Peeters, and P. Vasilopoulos, Phys. Rev. B
{\bf 76}, 115419 (2007). 

\bibitem{abergel}
D.S.L. Abergel and T. Chakraborty, Phys. Rev. Lett. {\bf 102}, 056807 (2009).

\bibitem{Falko_11}
M. Mucha-Kruczynski, I.L. Aleiner, and V.I. Fal'ko, Phys. Rev. B {\bf 84}, 
041404(R) (2011).

\bibitem{bi_rotate}
V.M. Apalkov and T. Chakraborty, Phys. Rev. B {\bf 84}, 033408 (2011).

\bibitem{latil}
S. Latil and L. Henrard, Phys. Rev. Lett. {\bf 97}, 036803 (2006);
S. Shallcross, S. Sharma, and O.A. Pankratov, Phys.
Rev. Lett. {\bf 101}, 056803 (2008).

\bibitem{lopes_07} J.M.B. Lopes dos Santos, N.M.R. Peres, and
A.H. Castro Neto, Phys. Rev. Lett. {\bf 99}, 256802 (2007).

\bibitem{mele_10} E.J. Mele, Phys. Rev B {\bf 81}, 161405 (2010).

\bibitem{bi_FQHE}
V.M. Apalkov and T. Chakraborty, Phys. Rev. Lett. {\bf 105}, 036801
(2010).

\bibitem{FQHE_spin_theory}
T. Chakraborty, Adv. Phys. {\bf 49}, 959 (2000); T. Chakraborty,
P. Pietil\"ainen, and F.C. Zhang, Phys. Rev. Lett. {\bf 57}, 130 (1986);
T. Chakraborty, Surf. Sci. {\bf 229}, 16 (1990); T. Chakraborty and
F.C. Zhang, Phys. Rev. B {\bf 29}, 7032 (1984); F.C. Zhang
and T. Chakraborty, {\it ibid.} {\bf 34}, 7076 (1986); T. Chakraborty and
P. Pietil\"ainen, {\it ibid.} {\bf 39}, 7971 (1989); {\bf 41}, 10862 (1990); 
Phys. Rev. Lett. {\bf 76}, 4018 (1996); {\bf 83}, 5559 (1999).

\bibitem{FQHE_spin_expt}
R.G. Clark, S.R. Haynes, A.M. Suckling, J.R. Mallett, P.A.
Wright, J.J. Harrish, and C.T. Foxon, Phys. Rev. Lett. {\bf 62}, 1536
(1989); J.P. Eisenstein, H.L. St\"ormer, L. Pfeiffer, and K.W. West,
{\it ibid.} {\bf 62}, 1540 (1989); A.G. Davis, R. Newbury, M. Pepper, 
J.E.F. Frost, D.A. Ritchie, and G.A.C. Jones, Phys. Rev. B {\bf 44}, 13128
(1991); L.W. Engel, S.W. Hwang, T. Sajoto, D.C. Tsui, and M.
Shayegan, {\it ibid.} {\bf 45}, 3418 (1992); T. Sajoto, Y.W. Suen, L.W. Engel,
M.B. Santos, and M. Shayegan, {\it ibid.} {\bf 41}, 8449 (1990).

\bibitem{tilted1} Y.-H. Hyun, Y. Kim, C. Sochichiu and
M.-Y. Choi, J. Phys.: Condens. Matter {\bf 24} 045501 (2012). 

\bibitem{igor}
I.V. Kukushkin, K. von Klitzing, and K. Eberl, Phys. Rev. B {\bf 55},
10607 (1997).

\bibitem{halperin} 
A. Stern and B.I. Halperin, Phys. Rev. Lett. {\bf 96}, 016802 (2006).

\bibitem{stern} A. Stern, Ann. Phys. {\bf 323}, 204 (2008).

\bibitem{majorana}
N. Read and D. Green, Phys. Rev. B {\bf 61}, 10267 (2000);
D.A. Ivanov, Phys. Rev. Lett. {\bf 86}, 268 (2001).

\bibitem{read}
N. Read, Physica {\bf B 298}, 121 (2001); G. Moore and N. Read,
Nucl. Phys. B {\bf 360}, 362 (1991).

\bibitem{anti1}
M. Levin, B.I. Halperin, and B. Rosenow, Phys. Rev. Lett. {\bf
99}, 236806 (2007).

\bibitem{anti2} S.S. Lee, S. Ryu, C. Nayak, and M.P.A. Fisher,
Phys. Rev. Lett. {\bf 99}, 236807 (2007).

\bibitem{greiter_2}
M. Greiter, X.-G. Wen, and F. Wilczek, Phys. Rev. Lett.
{\bf 66}, 3205 (1991); Nucl. Phys. {\bf B374}, 567 (1992).

\bibitem{Pfaffian_PRL} 
V. Apalkov and T. Chakraborty, Phys. Rev. Lett. {\bf 107}, 186803 (2011).

\bibitem{storni} M. Storni, R.H. Morf, S. Das Sarma,
Phys. Rev. Lett. {\bf 104}, 076803 (2010).

\end{thebibliography}
\end{document}